\documentclass[11pt]{article}

\usepackage[english]{babel}
\usepackage[latin1]{inputenc}
\usepackage{amsmath, latexsym}
\usepackage{amssymb}
\usepackage[authoryear]{natbib}
\usepackage{color}

\newif\ifpdf
\ifx\pdfoutput\undefined
   \pdffalse
\else
   \pdfoutput=1
   \pdftrue
\fi

\ifpdf
   \usepackage{graphicx}
    \definecolor{myred}{rgb}{0.5,0,0}
    \definecolor{myblue}{rgb}{0,0,0.75}
    \definecolor{mygreen}{rgb}{0,0.5,0}
    \usepackage[pdftex,
               pdftitle={Measuring sectoral diversification in an asymptotic multi-factor framework},
               pdfsubject={Asymptotic multi-factor model},
               pdfkeywords={credit portfolio model, capital requirement, quantile derivative},
               pdfauthor={Dirk Tasche},
               bookmarks=false,
               breaklinks=true,
               colorlinks=true,
               citecolor=myred,
               linkcolor=myblue,
               urlcolor=mygreen]{hyperref}
\else
   \usepackage{hyperref}
   \usepackage[dvips]{graphicx}
   \usepackage{rotating}
\fi

\newtheorem{theorem}{Theorem}[section]

\newtheorem{remark}[theorem]{Remark}
\newtheorem{example}[theorem]{Example}
\newtheorem{proposition}[theorem]{Proposition}
\newtheorem{definition}[theorem]{Definition}
\newtheorem{corollary}[theorem]{Corollary}
\newtheorem{assumption}[theorem]{Assumption}

\numberwithin{equation}{section}

\newlength{\captionwidth}
\begin{document}

\title{Measuring sectoral diversification in an asymptotic multi-factor framework}

\author{%
Dirk Tasche\thanks{Deutsche Bundesbank, Postfach 10 06 02, 60006
Frankfurt am Main, Germany\newline E-mail: dirk.tasche@gmx.net\newline
The opinions expressed in this paper are those of the
author and do not necessarily reflect views of Deutsche
Bundesbank. The author thanks Michael Gordy,  Katja Pluto, an
anonymous referee and participants in the workshop ``Concentration
Risk in Credit Portfolios'' (Eltville, November 2005) for useful
comments on an earlier version of the paper.}}

\date{July 2006}
\maketitle

\begin{abstract}
We investigate a multi-factor extension of the asymptotic single
risk factor (ASRF) model that underlies the capital charges
of the ``Basel II Accord''. In this extended model, it is still
possible to derive closed-form solutions for the risk
contributions to Value-at-Risk and Expected Shortfall. As an
application of the risk contribution formulae we introduce a new
concept for a diversification measure. The use of this new measure
is illustrated by an example calculated with a two-factor model.
The results with this model indicate that, thanks to dependence on
not fully correlated systematic sectors, there can be a
substantial reduction of risk contributions by sectoral
diversification effects.
\end{abstract}


\section{Introduction}

The Internal Ratings-Based Approach (IRBA) of Basel II \citep{BC04} is often
regarded as a first step towards supervisory recognition of portfolio credit risk
models for calculation of minimum capital requirements. IRBA capital formulae
were derived within a special portfolio credit model, the so-called ``Asymptotic
Single Risk Factor Model'' \citep[ASRF model,][]{Gordy03}. This model has the
property that capital allocations to individual positions depend only upon the
characteristics of these positions, but not upon the composition of the portfolio.
This property of ``portfolio invariance'' allows for computational simplicity, in
particular in that the required capital for a credit risk portfolio can be calculated
in a bottom-up approach by determining capital requirements at the position
level and adding them up. As a consequence, however, the model can reflect
neither name concentrations nor the sectoral structure of the portfolio (i.e., the
distribution of borrowers across industry and geographic regions).

The model's inability to capture name concentrations entails a
potential underestimation of the risk inherent in the portfolio,
whereas its fault in recognizing the potential diversification
effects following from the sectoral structure of the portfolio
could result in an overestimation of portfolio risk.
Overestimation of portfolio risk, however, can only occur if there
is no significant change in the assigned asset correlations
between the one-factor model and the multi-factor model that
reflects the sectoral structure. In comparison with a multi-factor
model where there are significantly higher average asset
correlations, the ASRF model even can underestimate portfolio
risk. As a consequence, whether the ASRF model underestimates the
risk of a certain portfolio or not has to be examined on a case by
case basis. The Basel Committee decided to deal in Pillar 2 of the
Basel II Accord \citep{BC04} with the assessment of name risk
concentration. As a consequence there is so far no automatism of
extended regulatory capital requirements for risk concentrations,
but banks will have to demonstrate to the supervisors that they
have established appropriate procedures to keep concentrations
under control. A quantitative way of tackling the name
concentration issue was suggested in \citet{EmmerTasche05}, for
instance.

In the present paper, we suggest a minimal -- in the spirit of
\citet{EmmerTasche05} -- extension of the Basel II model that
allows to study the effects of the sectoral structure on portfolio
risk. Admitting several risk factors instead of a single factor
only and applying the same transition to the limit as described in
\citet{Gordy03}, we arrive at versions of the model that remove
the  restriction of assuming a one-sector structure.
Alternatively, our class of models can be regarded as special
cases of the asymptotic models introduced by \citet[][Theorem
1]{Lucasetal01}.

As determining risk contributions or, economically speaking,
capital requirements for assets, sectors or sub-portfolios, is a
main purpose when using credit risk models, deriving exact
formulae for risk contributions to ``Value-at-risk'' (VaR) and
``Expected Shortfall'' (ES) in the asymptotic multi-risk factors
setting represents an important contribution of our paper to the
subject. Our results complement results on the differentiation of
VaR and ES presented in \citet{GLS00}, \citet{L99}, and
\cite{Tasche99}. In contrast to those papers, the results given
below can be applied for calculating economic capital or
contributions to economic capital without any further adaptation
of the formulae. From a computational point of view, the resulting
formulae are more demanding than in the one factor case, and --
necessarily, as otherwise diversification effects could not be
recognized -- the resulting risk contributions are not portfolio invariant any longer.

As an application of the risk contribution formulae we introduce
then a new concept for a measure of diversification, called
diversification factor. This factor can be computed at portfolio
as well as at sub-portfolio, sector or asset level, thus allowing
the main concentrations to be identified. The use of
these new factors is illustrated by an example calculated with a
two-factor model. The results with this model indicate that there
can be a substantial reduction of risk contributions by sectoral
diversification effects.

The material presented here is closely related to work by
\citet{Pykhtin04} and \citet{Garcia04}. \citeauthor{Pykhtin04}
describes an approximation of multi-factor models by single-factor
models, thus transferring the computational simplicity of
single-factor models to multi-factor models. \citeauthor{Garcia04}\
propose ``factor adjustments'' to stand-alone capital charges in
order to reflect diversification effects. As our results on the
risk contribution formulae are not approximate but exact they
could be used for benchmarking the results by
\citeauthor{Pykhtin04} and \citeauthor{Garcia04}

This paper is organized as follows: In Section \ref{se:basic} we
introduce the class of models we are going to analyze and derive
some basic properties. In Section \ref{se:computing} we briefly review
the Euler allocation principle that justifies the use of
partial derivatives as risk contributions and derive then the
formulae for risk contributions to VaR and ES in the
asymptotic multi-factor setting. A potential application of the
risk contribution formulae for the purpose of identifying sources
of risk concentration is suggested in Section \ref{se:defining}
where a new concept for a diversification measure is introduced.
Section \ref{se:example} gives a numerical illustration of a
potential application of the formulae and the diversification
factor. We conclude with some summarizing comments in Section
\ref{se:concl}.

\section{Asymptotic multi-factor models:
basic properties}
\label{se:basic}

The starting point for the factor models\footnote{%
See \citet{Bluhmetal} and the references therein for more
information on credit risk models. } we are going to consider is a
random variable $\tilde{L}(u)= \tilde{L}(u_1, \ldots, u_n)$ that
reflects the loss suffered from a portfolio of $n$ credit assets,
with respective exposures $u_i$. The tilde indicates that we
regard the original loss variable, without any approximation
procedure. The variable $\tilde{L}(u)$ can be interpreted as the
absolute loss, measured
in units of some currency. Then the $u_i$ are absolute
exposures\footnote{%
$u_i$ may also be thought as a face value multiplied with some
factor that expresses the average loss rate in case of default.}
and amounts of money. Alternatively, $\tilde{L}(u)$ can also be
understood as relative loss,
indicating the lost percentage of the sum of all exposures\footnote{%
If the underlying absolute loss variable incorporates average loss
rates in case of default, relative loss rather indicates the lost percentage
of the sum of all exposures, weighted with their average loss rates.%
}. In this case the $u_i$ are non-negative numbers without units
that add up to 1.

Formally, the original loss variable $\tilde{L}(u)$ is given as
\begin{equation}\label{eq:Ltilde}
\tilde{L}(u)\ =\ \sum_{i=1}^n u_i\,\mathbf{1}_{D_i}.
\end{equation}
The term $\mathbf{1}_{D_i}$ is the \emph{default indicator variable}
for asset $i$, i.e.\ it takes the value of 1 if $i$ defaults and $0$ if not.
As a consequence, the sum in \eqref{eq:Ltilde} will be built up with only those
$u_i$'s that relate to defaulted assets $i$. For factor models, it is
quite common to specify the default events $D_i$ by
\begin{equation}\label{eq:def_event}
D_i\ = \ \Big\{ \sum\nolimits_{j=1}^k \varrho_{i,j}\,S_j +
\omega_{i}\,\xi_i \le t_i\Big\}, \qquad i=1, \ldots, n,
\end{equation}
where the following holds for the involved constants and random
variables:
\begin{itemize}
    \item The random variables $S_1, \ldots, S_k$ are the \emph{systematic
risk factors}. They are assumed to capture the dependence of the default
events. In general, we have $k\ll n$. Within this paper, we assume
that the factor variables are standardized, i.e.
\begin{equation}\label{eq:stand}
\mathrm{E}[S_j] = 0\quad\text{and}\quad \mathrm{var}[S_j] = 1, \quad j=1, \ldots, k.
\end{equation}
The $S_1, \ldots, S_k$ may be stochastically dependent, but they do not have do be.
    \item The random variables $\xi_1, \ldots, \xi_n$ are the
    \emph{idiosyncratic risk drivers}. They
    are also standardized, i.e.
\begin{equation}
\mathrm{E}[\xi_i] = 0\quad\text{and}\quad \mathrm{var}[\xi_i] = 1, \quad i = 1, \ldots, n.
\end{equation}
$\xi_1, \ldots, \xi_n, (S_1, \ldots, S_k)$ are stochastically independent. As a consequence,
conditional on $(S_1, \ldots, S_k)$, the default events $D_i, i= 1,\ldots, n$ are independent.
    \item The constants $\varrho_{i,j}, i=1, \ldots,n,\,j= 1,
    \ldots, k$ are the \emph{factor loadings}
    of the systematic factors. We assume that
    \begin{equation}\label{eq:loadings}
\sum_{j=1, \ell=1}^k \varrho_{i,j}\,\varrho_{i,\ell}\,
\mathrm{corr}[S_j, S_\ell] \ \le \ 1, \quad i = 1, \ldots, n.
\end{equation}
By \eqref{eq:loadings} and the standardization assumption on the $S_j$ and $\xi_i$
the \emph{idiosyncratic loadings} $\omega_i, i= \ldots, n$ are well defined by
\begin{equation}\label{eq:omegas}
    \omega_i = \sqrt{1-\sum_{j=1, \ell=1}^k
\varrho_{i,j}\,\varrho_{i,\ell}\, \mathrm{corr}[S_j, S_\ell]}.
\end{equation}
As a further consequence of \eqref{eq:loadings} and
\eqref{eq:omegas} and of the standardization assumptions also the
\emph{asset values changes} $\sum\nolimits_{j=1}^k
\varrho_{i,j}\,S_j + \omega_{i}\,\xi_i$ are standardized.
\item The constant $t_i, i = 1, \ldots, n$ is called \emph{default threshold.}
It can be thought as a critical loss in value of borrower $i$'s assets that
causes the borrower to default on asset $i$. It is common to derive
$t_i$ from borrower $i$'s (assumed to be known) probability of default
$p_i$. Hence, we determine $t_i$ such that
\begin{equation}\label{eq:pi}
\mathrm{P}[D_i] \ =\ \mathrm{P}\Big[\sum\nolimits_{j=1}^k
\varrho_{i,j}\,S_j + \omega_{i}\,\xi_i \le t_i\Big]\ =\ p_i, \quad
i=1, \ldots n.
\end{equation}
When the idiosyncratic risk drivers $\xi_i$ and the factor
variables are all standard normally distributed, also the asset
value changes $\sum\nolimits_{j=1}^k \varrho_{i,j}\,S_j +
\omega_{i}\,\xi_i$ are standard normal. Let $\Phi$ denote the
standard normal distribution function. By \eqref{eq:pi} we then
have $p_i = \Phi(t_i)$.
\end{itemize}
Assuming conditional independence of the default events, given the
realizations of the systematic factors, entails by the law of
large numbers that the loss variable $\tilde{L}(u)$ can be
reasonably approximated by a modified loss variable $\hat{L}(u)$.
The approximate loss variable $\hat{L}(u)$ then depends on the
systematic factors only \citep[cf.][]{Gordy03, Lucasetal01}. This
is interpreted as elimination of the idiosyncratic risk by name
diversification. In general, the quality of the approximation
depends on conditions like the number of credit assets in the
portfolio, the granularity of the portfolio, or the correlations
of the asset value changes with the systematic factors.
$\hat{L}(u)$ is obtained from $\tilde{L}(u)$ by replacing the
default indicators $\mathbf{1}_{D_i}$ with their best predictors
given the systematic factors, i.e.\ with the conditional
probabilities $\mathrm{P}[D_i\,|\,(S_1, \ldots, S_k)]$. Hence
$\hat{L}(u)$ is given by
\begin{equation}\label{eq:limit_model}
\hat{L}(u)\ = \ \sum_{i=1}^n u_i\,\mathrm{P}[D_i\,|\,(S_1, \ldots,
S_k)].
\end{equation}
\begin{example}\label{ex:idio_phi}
If the default events are given by \eqref{eq:def_event} and the
idiosyncratic risk drivers $\xi_i$ are standard normally
distributed, then the approximate loss variable $\hat{L}(u)$ can
be written as
\begin{equation}\label{eq:normal_xi}
    \hat{L}(u)\ = \ \sum_{i=1}^n u_i\,\Phi\bigg(
    \frac{t_i -\sum\nolimits_{j=1}^k \varrho_{i,j}\,S_j}{\omega_i}\bigg).
\end{equation}
\end{example}
Example \ref{ex:idio_phi} suggests to consider a --
compared to \eqref{eq:limit_model} -- slightly \emph{generalized loss
variable} $L(u)$
\begin{equation}\label{eq:new_model}
    L(u) \ = \ \sum_{i=1}^n u_i\,g_i(S) \ =\ \sum_{i=1}^n u_i\,g_i(S_1, \ldots,
    S_k),
\end{equation}
with $g_i: \mathbb{R}^k \to [0,1]$, $i = 1, \ldots, n$ decreasing\footnote{%
The results of this paper hold also when ``decreasing'' is
replaced by ``increasing''. Some of the formulae then must be
appropriately adapted. } at least in one (always the same)
component of the vector argument. Considering the loss variable in
this generalized form might be of interest, in particular, when
losses are measured mark-to-market.

When investigating the generalized loss variable $L(u)$ we will
need some technical conditions and notations as specified in the
following assumption and \eqref{eq:Stilde},
\eqref{eq:distribution} and \eqref{eq:G1}.
\begin{assumption}\label{as:technical}\
\vspace{-2ex}
\begin{enumerate}
    \item The exposures $u_i, i = 1, \ldots,n $ in definition \eqref{eq:new_model}
    are non-negative.
    \item For any fixed $(k-1)$-tuple $(s_2, \ldots, s_k)$ the mapping
\begin{equation}\label{eq:mapping}
    s_1 \ \mapsto \ \sum_{i=1}^n u_i\,g_i(s_1, \ldots, s_k),
    \quad \mathbb{R}\to [0,\infty[
\end{equation}
is strictly decreasing, continuous, and onto
$]0,\,\mathcal{U}\,[$, with $\mathcal{U}$ defined by $\mathcal{U}
= \sum_{i=1}^n u_i$.
\item There is a conditional density $h(s_1\,|\,s_2, \ldots, s_k)$ of
$S_1$ given $(S_2, \ldots, S_k)$.
\end{enumerate}
\end{assumption}
The condition that the mapping from \eqref{eq:mapping} is
onto $]0,\,\mathcal{U}\,[$ is, in particular, satisfied when
$L(u)$ is specified as in Example \ref{ex:idio_phi} with
$\varrho_{i,1} >0$ for all $i = 1, \ldots, n$. The condition,
however, could be dispensed with, at the price of having more
complexity in the following results on densities and risk
contributions.
For the sake of a more concise notation we define
\begin{subequations}
\begin{equation}\label{eq:Stilde}
    \widetilde{S}\ =\ (S_2, \ldots, S_k),\qquad \widetilde{s}\ =\
    (s_2, \ldots, s_k).
\end{equation}
The distribution $\mathrm{P}\widetilde{S}^{-1}$ of $\widetilde{S}$
is then given by the relation
\begin{equation}\label{eq:distribution}
 \mathrm{P}\widetilde{S}^{-1}(A) \ = \ \mathrm{P}[\widetilde{S} \in A]
\end{equation}
that holds for all Borel-sets $A \subset \mathbb{R}^{k-1}$. In
particular, if $\widetilde{S}$ has a density $\phi$, then
integration with respect to the distribution of $\widetilde{S}$
can be expressed as $\mathrm{P}\widetilde{S}^{-1}(d \widetilde{s})
= \phi(\widetilde{s})\,d \widetilde{s}$. For fixed $u$ write
\begin{equation}\label{eq:G1}
    G(v, \widetilde{s})\ =\ \sum_{j=1}^n
    u_j\,g_j(v,\widetilde{s}).
\end{equation}
\end{subequations}
By Assumption \ref{as:technical}, then, for fixed $\widetilde{s}$,
$v \mapsto G(v, \widetilde{s})$ is invertible. Write
$G(\cdot,\widetilde{s})^{-1}$ for the inverse function of $v
\mapsto G(v, \widetilde{s})$. Write additionally $G(\cdot,
\widetilde{s})^{-1}(0) = \infty$ and $G(\cdot,
\widetilde{s})^{-1}(z) = -\infty$ for $z \ge \mathcal{U}$.

Having fixed the assumptions and notations, we can prove a result
on the calculation of moments that in particular implies that the
distribution of the generalized loss variable $L(u)$ has a
density. Moreover, Proposition \ref{pr:1} will be applied, in the
context of model \eqref{eq:new_model}, for identifying the VaR
contributions as conditional expectations.
\begin{proposition}\label{pr:1}
Let $F: [0,1] \to\mathbb{R}$ be arbitrary and $L(u)$ be the
loss variable defined by \eqref{eq:new_model}. Then, under Assumption\footnote{%
In order to keep the representation of the results as intuitive
and clear as possible here and in the following the proofs will
not be rigorous but rather consist of calculations without
consideration of continuity, differentiability etc. Moreover, for
some of the results, additional assumptions on existence of
moments etc. must be made. } \ref{as:technical}, for any $0 \le
z\le \mathcal{U}$ and $i\in\{1, \ldots,n\}$ we have:
\begin{equation*}
    \mathrm{E}\bigl[F(g_i(S))\,\mathbf{1}_{\{L(u) \le z\}}\bigr]
    =
    - \int_0^z \mathrm{E}\bigg[
    \frac{F\bigl(g_i(G(\cdot, \widetilde{S})^{-1}(t), \widetilde{S})\bigr)\,
    h\bigl(G(\cdot, \widetilde{S})^{-1}(t)\,|\,\widetilde{S}\bigr)}
    {\frac{\partial}{\partial v}G(v, \widetilde{S})\bigl|_{v=G(\cdot,
    \widetilde{S})^{-1}(t)}}\bigg]\,d t.
\end{equation*}
\end{proposition}
\textbf{Proof.}
\begin{subequations}
\begin{align}
    \mathrm{E}\bigl[F(g_i(S))\,\mathbf{1}_{\{L(u) \le z\}}\bigr] &=
    \int \mathrm{E}\bigl[F(g_i(S_1, \widetilde{S}))\,
    \mathbf{1}_{\{G(S_1, \widetilde{S}) \le z\}}\,|
    \,\widetilde{S}=\widetilde{s}\bigr]
    \mathrm{P}\widetilde{S}^{-1}(d \widetilde{s})\label{eq:fundamental}\\
\intertext{(taking into account $G(\cdot,
\widetilde{s})^{-1}(0) = \infty$)} &=
\int \int_{G(\cdot, \widetilde{s})^{-1}(z)}^\infty
F\bigl(g_i(y, \widetilde{s})\bigr)\,h(y\,|\,\widetilde{s})\,d y
\,\mathrm{P}\widetilde{S}^{-1}(d \widetilde{s})\label{eq:disint}\\
\intertext{(substituting $y = G(\cdot, \widetilde{s})^{-1}(t)$)}
&=  - \int \int_0^z
    \frac{F\bigl(g_i(G(\cdot, \widetilde{s})^{-1}(t), \widetilde{s})\bigr)\,
    h\bigl(G(\cdot, \widetilde{s})^{-1}(t)\,|\,\widetilde{s}\bigr)}
    {\frac{\partial}{\partial v}G(v, \widetilde{S})\bigl|_{v=G(\cdot,
    \widetilde{S})^{-1}(t)}}
    \,d t \,\mathrm{P}\widetilde{S}^{-1}(d \widetilde{s}).
    \notag
\end{align}
\end{subequations}
The assertion follows by applying Fubini's theorem.\hfill $\Box$

The choice $F=1$ in Proposition \ref{pr:1} implies the existence of
a density for the distribution of the generalized
loss variable:
\begin{corollary}\label{co:1}
Under Assumption \ref{as:technical}, the loss variable $L(u)$ from
\eqref{eq:new_model} has the density $f_{L(u)}: ]0,\mathcal{U}[
 \to [0,\infty[$,
defined by
\begin{equation}\label{eq:density}
    f_{L(u)}(t) \ =\ - \mathrm{E}\left[
    \frac{
    h\bigl(G(\cdot, \widetilde{S})^{-1}(t)\,|\,\widetilde{S}\bigr)}
    {\frac{\partial}{\partial v}G(v, \widetilde{S})\bigl|_{v=G(\cdot,
    \widetilde{S})^{-1}(t)}}\right], \quad t \in
    \,]0,\,\mathcal{U}\,[.
\end{equation}
\end{corollary}
Note that \citeauthor{Gordy03}'s (\citeyear{Gordy03}) ASRF (Asymptotic Single Risk Factor) model is
a special case of \eqref{eq:limit_model} with $k=1$. Then the expectation in \eqref{eq:density}
disappears, and the density $h$ and the inverse of $G$ do not depend on $\widetilde{s}$.
Nevertheless, in the case of non-constant asset correlations or probabilities of default
the calculation of the density of $L(u)$ will involve numerical inversion of $G$ even in
the simple ASRF case.

In case that a closed-form representation of the conditional
distribution of $S_1$ given $\widetilde{S}$ is known (e.g.\ if $S$
is jointly normally distributed), Equation \eqref{eq:fundamental}
(with $F=1$) immediately yields a more efficient way to calculate
the distribution function of $L(u)$ than Corollary \ref{co:1}
does. The reason is that application of Corollary \ref{co:1} would
require evaluation of a $k$-dimensional integral if $S$ had a
density, whereas the application of the following Proposition
\ref{pr:2} would only require evaluation of a $(k-1)$-dimensional
integral.
\begin{proposition}\label{pr:2}
Under Assumption \ref{as:technical}, the distribution function of
the loss variable $L(u)$ as given in \eqref{eq:new_model} can be calculated by means of
\begin{equation}\label{eq:distr_funct}
    \mathrm{P}[L(u)\le z] \ = \ \int \mathrm{P}\bigl[S_1 \ge G(\cdot, \widetilde{S})^{-1}(z)\,|\,
    \widetilde{S}=\widetilde{s}\bigr]\,\mathrm{P}\widetilde{S}^{-1}(d \widetilde{s}).
\end{equation}
\end{proposition}
Define, for $\alpha \in (0,1)$ and any real random variable $X$, the $\alpha$-\emph{quantile}
of $X$ by
\begin{equation}\label{eq:quantile}
    q_\alpha(X) \ = \ \min\{x: \mathrm{P}[X \le x]\ge \alpha\}.
\end{equation}
Quantiles at high levels (e.g.\ 99.9\%) are popular metrics for
determining the economic capital of portfolios. Within the
financial community, the $\alpha$-quantile of a loss distribution
is commonly called \emph{Value-at-Risk} (VaR) at level $\alpha$.
In case of the generalized loss variable $L(u)$ the quantiles
$q_\alpha(L(u))$ can be computed by numerical inversion of
\eqref{eq:distr_funct}.

We conclude this section by providing two alternative formulae for the calculation
of another popular risk measure, the Expected Shortfall\footnote{%
See \citet{AcerbiTasche} and the references given therein for more
details on Expected Shortfall vs.\ Value-at-Risk. In particular,
in case of discontinuous loss distributions the definition of
ES has to be slightly modified in order to make it a risk measure superior to VaR.%
}, in the case of the asymptotic
multi-factor model under consideration.
\begin{remark}\label{rm:ES}
The Expected Shortfall
$\mathrm{ES}_\alpha(L(u)) = \mathrm{E}[L(u)\,|\,L(u) \ge q_\alpha(L(u))]$
at level $\alpha$
of the loss variable $L(u)$ from \eqref{eq:new_model} can alternatively
be calculated with recourse to Corollary \ref{co:1} or to Proposition \ref{pr:2}.
Note that  the existence of a
density of the distribution of $L(u)$ (Corollary \ref{co:1})
implies $\mathrm{P}[L(u) \ge q_\alpha(L(u))] = 1-\alpha$.
From Corollary \ref{co:1} we can therefore derive
\begin{subequations}
\begin{equation}\label{eq:ESa}
    \mathrm{E}[L(u)\,|\,L(u) \ge q_\alpha(L(u))] \ =\ - (1-\alpha)^{-1}
    \int_{q_\alpha(L(u))}^{\mathcal{U}}
    t\,\mathrm{E}\left[
    \frac{
    h\bigl(G(\cdot, \widetilde{S})^{-1}(t)\,|\,\widetilde{S}\bigr)}
    {\frac{\partial}{\partial v}G(v, \widetilde{S})\bigl|_{v=G(\cdot,
    \widetilde{S})^{-1}(t)}}\right] d t.
\end{equation}
From Proposition \ref{pr:2} we obtain (by making use of the formula $\mathrm{E}[X] =
\int_0^\infty \mathrm{P}[X\ge x]\, d x$ for $X \ge 0$)
\begin{eqnarray}
\lefteqn{\mathrm{E}[L(u)\,|\,L(u) \ge q_\alpha(L(u))]}\label{eq:ESb}\\
    &= &q_\alpha(L(u))
    +
    (1-\alpha)^{-1}\,
    \int\limits_{q_\alpha(L(u))}^{\mathcal{U}} \int
    \mathrm{P}\bigl[S_1 \le G(\cdot, \widetilde{S})^{-1}(z)\,|\,
    \widetilde{S}=\widetilde{s}\bigr]\,\mathrm{P}\widetilde{S}^{-1}(d \widetilde{s})\, d z.
    \notag
\end{eqnarray}
\end{subequations}
\end{remark}

\section{Computing the risk contributions}
\label{se:computing}

When economic capital for a portfolio is determined by means of a
homogeneous risk measure, according to the Euler allocation
principle -- to be introduced in Section \ref{se:euler} -- the
risk contributions of assets should be calculated as partial
derivatives of the portfolio-wide economic capital with respect to
the exposures. In Section \ref{se:partial} we will then derive
formulae for the derivatives
of Value-at-Risk\footnote{%
See \citet{Mausser04} and the references therein for
the practical issues when estimating VaR contributions
from statistical samples.%
} as defined by \eqref{eq:quantile} and Expected
Shortfall as defined in Remark \ref{rm:ES} in the context of
the asymptotic multi-factor model of Section
\ref{se:basic}.

\subsection{Euler allocation}
\label{se:euler}

Suppose that real-valued random variables $X_1, \ldots, X_n$ are given that
stand for the profits and losses with the assets in a portfolio. Let $Y$
denote the portfolio-wide profit and loss, i.e.\ let
\begin{equation}\label{eq:defY}
    Y\ =\ \sum_{i=1}^n X_i.
\end{equation}
The economic capital $\mathrm{EC}$ required by the portfolio is
determined with a risk measure $\rho$, i.e.\
\begin{equation}\label{eq:ec}
    \mathrm{EC}\ =\ \rho(Y).
\end{equation}
\begin{definition}\label{de:risk_contrib}
If $\rho$ is a risk measure and $V, W$ are random variables such
that the derivative $\frac{d}{d h}\rho(h\,V+W)\bigl|_{h=0}$
exists, then
\begin{equation*}
    \rho(V\,|\,W)\ =\ \frac{d}{d h}\rho(h\,V+W)\bigl|_{h=0}
\end{equation*}
is called \emph{contribution of $V$ to the risk of $W$ in respect
of $\rho$}.
\end{definition}
It is natural to require that, in a portfolio with loss variable
as in \eqref{eq:defY}, the risk contributions add up to the
portfolio EC. As we will see below, this property is closely
related to a homogeneity property of the risk measure $\rho$.
\begin{assumption}\label{as:euler}
The risk measure $\rho$ is positively homogeneous, i.e.
\begin{equation*}
    \rho(h\,Z) \ =\ h\,\rho(Z)
\end{equation*}
for any random variable $Z$ in the definition set of $\rho$ and $h
> 0$.
\end{assumption}
This assumption seems very natural as long as the asset or
    portfolio under
    consideration is not significant compared to the market as
    a whole and is not subject
    to market liquidity risk.

If for every $i$ the contribution of $X_i$ to the risk of $Y$ exists, then
we have by Euler's theorem on the representation of positively homogenous functions
\begin{equation}\label{eq:euler}
    \rho(Y) \ =\ \sum_{i=1}^n \rho(X_i\,|\,Y).
\end{equation}
Hence, as required, positive homogeneity implies
    that, within a portfolio, the risk contributions add up to the total risk. This
    additivity property is of high practical importance.

The decomposition of the portfolio risk $\rho$ as given by
\eqref{eq:euler} is called \emph{Euler allocation}. The use of the
Euler allocation principle was justified by several authors with
different reasonings:
\begin{itemize}
    \item \citet{Patriketal} argued from a practitioner's view emphasizing
    mainly the fact that the risk contributions according to the Euler principle
    by \eqref{eq:euler} naturally add up to the portfolio-wide economic capital.
    \item \citet{Litterman96} and \citet{Tasche99} pointed out that the Euler
    principle is fully compatible with economically sensible portfolio diagnostics
    and optimization.
    \item \citet{Denault01} derived the Euler principle
    by game-theoretic considerations.
    \item In the context of capital allocation for insurance companies, \cite{MyersRead}
    argued that applying the Euler principle to the expected ``default value'' (essentially
    $\mathrm{E}[\max(Y, 0)]$) of the insurance portfolio is most appropriate for deriving
    line-by-line surplus requirements.
    \item More recently \citet{Kalkbrener05} presented an axiomatic approach to
    capital allocation and risk contributions. One of his axioms requires
    that risk contributions do not exceed the corresponding stand-alone risks.
    From this axiom in connection with more technical conditions, in the context
    of sub-additive and positively homogeneous risk measures, \citeauthor{Kalkbrener05}
    concluded that the Euler principle is the only allocation principle to be
    compatible with the ``diversification''-axiom \citep[see also][]{Kalkbreneretal04, Tasche02}.
\end{itemize}

\subsection{Partial derivatives of VaR and ES}
\label{se:partial}

Before coming to the main result on the partial derivatives
of VaR with respect to the exposures of the assets in the portfolio,
we will shortly discuss the case
of \citeauthor{Gordy03}'s (\citeyear{Gordy03}) ASRF model.
\begin{example}\label{ex:asrf}
In \citeauthor{Gordy03}'s (\citeyear[][cf.\ Proposition
4]{Gordy03}) ASRF model Equation \eqref{eq:new_model} reads
\begin{equation}\label{eq:gordy}
    L(u) \ = \ \sum_{j=1}^n u_j\,g_j(S),
\end{equation}
where $S$, the single systematic factor, stands for a
 random variable that satisfies some additional
conditions. The $g_j$ are strictly increasing and continuous
functions. As a consequence, the terms $g_j(S)$ in
\eqref{eq:gordy} are comonotonic (see Section \ref{se:defining}
for a definition) random variables. From the comonotonic
additivity of VaR and ES \citep[see, e.g.,][]{Tasche02} follows
then for any $\alpha \in (0,1)$ that
\begin{subequations}
\begin{align}\label{eq:comonotonicVaR}
     q_\alpha(L(u)) &=
     \sum_{j=1}^n u_j\,q_\alpha(g_j(S))\\
    \intertext{and}
    \mathrm{E}[L(u)\,|\,L(u) \ge q_\alpha(L(u))] &=
    \sum_{j=1}^n u_j\,\mathrm{E}[g_j(S)
    \,|\,g_j(X) \ge q_\alpha(g_j(S))].
    \label{eq:comonotonicES}
\end{align}
\end{subequations}
As the right-hand sides of \eqref{eq:comonotonicVaR} and \eqref{eq:comonotonicES}
are linear in the exposure vector $u$, applying the Euler allocation principle
with partial derivatives with respect to the components of $u$ yields
that the risk contributions to VaR or ES in the ASRF model model equal the corresponding
stand-alone risks.
\end{example}
In the following, we compute the derivatives of VaR and ES in the context
of an asymptotic multi-factor model as given by \eqref{eq:new_model}.
The validity of the results is subject to technical
conditions similar to those of \citet[][Section 5]{Tasche99}.
For reasons of readability of the text
we do not discuss these conditions here in detail.

\begin{subequations}
 Write
(slightly modifying the notation from \eqref{eq:G1} but keeping
\eqref{eq:Stilde})
\begin{align}\label{eq:G_ext}
    G(v, \widetilde{s}, u) &=
    \sum_{j=1}^n u_j\,g_j(v, \widetilde{s})\\
    \intertext{as well as}
    G_{(\widetilde{s}, u)}^{-1}(z) &
    = G(\cdot, \widetilde{s}, u)^{-1}(z)
    \label{eq:Ginv}.
\end{align}
Hence $G_{(\widetilde{s}, u)}^{-1}(z)$ denotes the solution $v^\ast$
of the equation
\begin{equation}\label{eq:solution}
    G(v^\ast, \widetilde{s}, u) \ =\ z
\end{equation}
with fixed $\widetilde{s}$ and $u$. Note that existence and
uniqueness of $G_{(\widetilde{s}, u)}^{-1}(z)$ is guaranteed by
Assumption \ref{as:technical}. Thus prepared, we can state the
main result of this paper (Theorem \ref{pr:3}), namely that in the
asymptotic multi-factor model the risk contributions to VaR,
calculated as partial derivatives, coincide with certain
expectations conditional on the portfolio loss equalling VaR.
\end{subequations}
\begin{theorem}\label{pr:3}
Under Assumption \ref{as:technical}\footnote{%
Some further technical conditions on uniform integrability of the
random variables under consideration have to be required, cf.\
\citet[][Lemma 5.3]{Tasche99} for a similar result. }, the
quantiles (VaRs) $q_\alpha(u)=q_\alpha(L(u))$ at level $\alpha\in
(0,1)$ of the generalized loss variable $L(u)$ as defined in
\eqref{eq:new_model} are partially differentiable with respect to
the portfolio weights $u_i$ of the single loss variables. The
partial derivatives $\frac{\partial q_\alpha(u)} {\partial u_i}$
are given by
\begin{equation}\label{eq:partial_deriv}
        \frac{\partial q_\alpha(u)}
{\partial u_i}\ =\ \mathrm{E}\bigg[
    \frac{
    h\bigl(G_{(\widetilde{S}, u)}^{-1}(q_\alpha(u))\,\bigr|\,\widetilde{S}\bigr)}
    {\frac{\partial}{\partial v} G\bigl(v,
    \widetilde{S},u\bigr)\bigl|_{v=G^{-1}_{(\widetilde{S},u)}(q_\alpha(u))}}\bigg]^{-1}
    \mathrm{E}\bigg[
    \frac{g_i\bigl(G_{(\widetilde{S}, u)}^{-1}(q_\alpha(u)), \widetilde{S},u\bigr)\,
    h\bigl(G_{(\widetilde{S}, u)}^{-1}(q_\alpha(u))\,\bigr|\,\widetilde{S}\bigr)}
    {\frac{\partial}{\partial v} G\bigl(v,
    \widetilde{S},u\bigr)\bigl|_{v=G^{-1}_{(\widetilde{S},u)}(q_\alpha(u))}}\bigg].
\end{equation}
\end{theorem}
\textbf{Proof.} Fix $z\,\in]0,\,\mathcal{U}\,[$ and observe from
\eqref{eq:solution} that
\begin{equation*}
    G\bigl(G^{-1}_{(\widetilde{s},u)}(z), \widetilde{s}, u\bigr)\ =\
    z \quad\text{for all}\ \widetilde{s},u
\end{equation*}
implies by \eqref{eq:G_ext}
\begin{align*}
    0 &= \frac{\partial}{\partial u_i}G\bigl(
    G^{-1}_{(\widetilde{s},u)}(z), \widetilde{s}, u)\bigr)\\
    &= \frac{\partial}{\partial u_i}G^{-1}_{(\widetilde{s},u)}(z) \frac{\partial}{\partial v} G\bigl(v,
    \widetilde{s},u\bigr)\bigl|_{v=G^{-1}_{(\widetilde{s},u)}(z)}
    +\, \frac{\partial}{\partial w_i} G\bigl(G^{-1}_{(\widetilde{s},u)}(z),
    \widetilde{s},w\bigr)\bigl|_{w=u}\\
    &=\frac{\partial}{\partial u_i}G^{-1}_{(\widetilde{s},u)}(z) \frac{\partial}{\partial v} G\bigl(v,
    \widetilde{s},u\bigr)\bigl|_{v=G^{-1}_{(\widetilde{s},u)}(z)}
    +\, g_i\bigl(G^{-1}_{(\widetilde{s},u)}(z),
    \widetilde{s}\bigr)
\end{align*}
and as a further consequence
\begin{subequations}
\begin{equation}\label{eq:implicit}
\frac{\partial}{\partial u_i}G^{-1}_{(\widetilde{s},u)}(z) \ =\
- \frac{g_i\bigl(G^{-1}_{(\widetilde{s},u)}(z),
    \widetilde{s}\bigr)}{\frac{\partial}{\partial v} G\bigl(v,
    \widetilde{s},u\bigr)\bigl|_{v=G^{-1}_{(\widetilde{s},u)}(z)}}.
\end{equation}
Additionally, we have
\begin{equation}\label{eq:inv_deriv}
    \frac{\partial}{\partial z} G^{-1}_{(\widetilde{s},u)}(z)\ =\ \left(
    \frac{\partial}{\partial v} G\bigl(v,
    \widetilde{s},u\bigr)\bigl|_{v=G^{-1}_{(\widetilde{s},u)}(z)}\right)^{-1}.
\end{equation}
\end{subequations}
Assuming existence\footnote{%
Under appropriate smoothness and moment conditions, existence can
be proven by means of the implicit function theorem. } of
$\frac{\partial q_\alpha(u)} {\partial u_i}$, it can implicitly be
determined as follows:
\begin{align}
    \alpha &= \mathrm{P}[L(u) \le q_\alpha(u)]\notag\\
&= \mathrm{E}\bigl[\mathrm{P}[S_1 \ge
G_{(\widetilde{S},u)}^{-1}(q_\alpha(u))\,|\,\widetilde{S}]\bigr]\notag\\
&= \mathrm{E}\left[\int_{G_{(\widetilde{S},u)}^{-1}(q_\alpha(u))}^\infty
h(y\,|\,\widetilde{S})\,d y \right]\notag\\
\intertext{implies}
0 &= - \mathrm{E}\left[\frac{\partial}{\partial u_i}G_{(\widetilde{S},u)}^{-1}
(q_\alpha(u))\,h\bigl(G_{(\widetilde{S},u)}^{-1}
(q_\alpha(u))\,|\,\widetilde{S}\bigr)\right].\label{eq:first_step}
    \end{align}
By \eqref{eq:implicit} and \eqref{eq:inv_deriv} we obtain
\begin{align}
    \frac{\partial}{\partial u_i}G_{(\widetilde{S},u)}^{-1}(q_\alpha(u))
    &= \frac{\partial}{\partial u_i}G^{-1}_{(\widetilde{S},u)}(z)\bigl|_{z=q_\alpha(u)}
    + \frac{\partial}{\partial z} G^{-1}_{(\widetilde{S},u)}(z)\bigl|_{z=q_\alpha(u)}
        \frac{\partial q_\alpha(u)}{\partial u_i}\notag\\
        &=\left(
    \frac{\partial}{\partial v} G\bigl(v,
    \widetilde{S},u\bigr)\bigl|_{v=G^{-1}_{(\widetilde{S},u)}(q_\alpha(u))}\right)^{-1}
    \left(\frac{\partial q_\alpha(u)}{\partial u_i}-
    g_i\bigl(G^{-1}_{(\widetilde{S},u)}(q_\alpha(u)),
    \widetilde{S}\bigr)\right).\label{eq:right_hand}
\end{align}
Replacing $\frac{\partial}{\partial u_i}G_{(\widetilde{S},u)}^{-1}(q_\alpha(u))$ in
\eqref{eq:first_step} by the right-hand side of \eqref{eq:right_hand} and solving for
$\frac{\partial q_\alpha(u)}{\partial u_i}$ yields the
assertion.\hfill $\Box$
\begin{remark}\label{rm:cond_exp}
Equation \eqref{eq:partial_deriv} may equivalently be written as
\begin{equation}\label{eq:cond_exp}
    \frac{\partial q_\alpha(L(u))}
{\partial u_i}\ =\ \mathrm{E}[g_i(S)\,|\,L(u)=q_\alpha(L(u))].
\end{equation}
This follows from Proposition \ref{pr:1} and Corollary \ref{co:1}.
For by Corollary \ref{co:1}, we have for any
$z\,\in]0,\,\mathcal{U}\,[$
\begin{subequations}
\begin{equation}\label{eq:denominator}
    \mathrm{E}\bigl[g_i(S)\,\mathbf{1}_{\{L(u) \le z\}}\bigr] \ = \
- \int_0^z \mathrm{E}[g_i(S)\,|\,L(u)=t]\, \mathrm{E}\left[
    \frac{
    h\bigl(G(\cdot, \widetilde{S})^{-1}(t)\,|\,\widetilde{S}\bigr)}
    {\frac{\partial}{\partial v}G(v, \widetilde{S})\bigl|_{v=G(\cdot,
    \widetilde{S})^{-1}(t)}}\right] d t.
\end{equation}
On the other hand, Proposition \ref{pr:1} implies with $F(x)=x$
that
\begin{equation}\label{eq:numerator}
\mathrm{E}\bigl[g_i(S)\,\mathbf{1}_{\{L(u) \le z\}}\bigr] \ =\
- \int_0^z \mathrm{E}\left[
    \frac{g_i(G(\cdot, \widetilde{S})^{-1}(t), \widetilde{S})\,
    h\bigl(G(\cdot, \widetilde{S})^{-1}(t)\,|\,\widetilde{S}\bigr)}
    {\frac{\partial}{\partial v}G(v, \widetilde{S})\bigl|_{v=G(\cdot,
    \widetilde{S})^{-1}(t)}}\right]\,d t.
\end{equation}
\end{subequations}
Equating the right-hand sides of \eqref{eq:denominator} and
\eqref{eq:numerator} respectively implies by taking the derivative
with respect to $z$ and then letting $z=q_\alpha(L(u))$ that
$\mathrm{E}[g_i(S)\,|\,L(u)=q_\alpha(L(u))]$ equals the
right-hand side of \eqref{eq:partial_deriv}.\hfill $\Box$
\end{remark}
A result analogous to Theorem \ref{pr:3} for VaR holds for ES as the following
corollary shows.
\begin{corollary}\label{co:ES}
Under Assumption \ref{as:technical}, the Expected Shortfall risk
measure $\mathrm{E}[L(u)\,|\,L(u)\ge q_\alpha(u)]$ (with
$q_\alpha(u)=q_\alpha(L(u))$) of the generalized loss variable as
defined in \eqref{eq:new_model} is partially differentiable with
respect to the weights $u_i$. The partial derivatives can be
computed as
\begin{equation}\label{eq:ES_deriv}
    \frac{\partial}{\partial u_i}\mathrm{E}[L(u)\,|\,L(u)\ge q_\alpha(u)]\ =\
    \mathrm{E}[g_i(S)\,|\,L(u)\ge q_\alpha(u)], \quad i = 1,
    \ldots, n.
\end{equation}
\end{corollary}
\textbf{Proof.} A straight-forward calculation as in the proof of
Proposition \ref{pr:1} (see \eqref{eq:disint}) yields
\begin{eqnarray}
    \lefteqn{(1-\alpha)\,\frac{\partial}{\partial u_i}\mathrm{E}[L(u)\,|\,L(u)\ge
    q_\alpha(u)]}\notag\\
    & = &
    \frac{\partial}{\partial u_i} \left(\sum_{j=1}^n u_j\,
    \mathrm{E}\left[\int_{-\infty}^{G^{-1}_{(\widetilde{S},u)}(q_\alpha(u))}
    g_j(y, \widetilde{S})\,h(y\,|\,\widetilde{S})\,d y\right]\right)\notag\\
    &=&\sum_{j=1}^n u_j\,
    \mathrm{E}\left[ \frac{\partial}{\partial u_i}
    G^{-1}_{(\widetilde{S},u)}(q_\alpha(u))\,
    g_j\bigl(G^{-1}_{(\widetilde{S},u)}(q_\alpha(u)),
    \widetilde{S}\bigr)\,
    h\bigl(G^{-1}_{(\widetilde{S},u)}(q_\alpha(u))\,\bigl|\,\widetilde{S}\bigr)\right]\notag\\
    & & \quad + \,\mathrm{E}\bigl[g_i(S)\,\mathbf{1}_{\{L(u)\ge q_\alpha(u)\}}\bigr].
    \label{eq:insert}
\end{eqnarray}
Making use of identity \eqref{eq:right_hand} and of the definition of
$G^{-1}_{(\widetilde{S},u)}$ (see \eqref{eq:G_ext} and \eqref{eq:Ginv}) we obtain
\begin{eqnarray*}
\lefteqn{\sum_{j=1}^n u_j\,\mathrm{E}\left[ \frac{\partial}{\partial u_i}
    G^{-1}_{(\widetilde{S},u)}(q_\alpha(u))\,
    g_j\bigl(G^{-1}_{(\widetilde{S},u)}(q_\alpha(u)),
    \widetilde{S}\bigr)\,
    h\bigl(G^{-1}_{(\widetilde{S},u)}(q_\alpha(u))\,\bigl|\,\widetilde{S}\bigr)\right]}\notag\\
  &=&  \frac{\partial q_\alpha(u)}{\partial u_i}\,
  \mathrm{E}\left[
    \frac{\left(\sum_{j=1}^n u_j\,g_j\bigl(G^{-1}_{(\widetilde{S},u)}(q_\alpha(u)),
    \widetilde{S}\bigr)\right)
    h\bigl(G_{(\widetilde{S}, u)}^{-1}(q_\alpha(u))\,\bigr|\,\widetilde{S}\bigr)}
    {\frac{\partial}{\partial v} G\bigl(v,
    \widetilde{S},u\bigr)\bigl|_{v=G^{-1}_{(\widetilde{S},u)}(q_\alpha(u))}}\right]\\
   & & \quad - \,\mathrm{E}\left[
    \frac{\left(\sum_{j=1}^n u_j\,g_j\bigl(G^{-1}_{(\widetilde{S},u)}(q_\alpha(u)),
    \widetilde{S}\bigr)\right) g_i\bigl(G^{-1}_{(\widetilde{S},u)}(q_\alpha(u)),
    \widetilde{S}\bigr)
    h\bigl(G_{(\widetilde{S}, u)}^{-1}(q_\alpha(u))\,\bigr|\,\widetilde{S}\bigr)}
    {\frac{\partial}{\partial v} G\bigl(v,
    \widetilde{S},u\bigr)\bigl|_{v=G^{-1}_{(\widetilde{S},u)}(q_\alpha(u))}}\right]\\
   &=& q_\alpha(u) \Bigg\{\frac{\partial q_\alpha(u)}{\partial u_i}\,\mathrm{E}\left[
    \frac{
    h\bigl(G_{(\widetilde{S}, u)}^{-1}(q_\alpha(u))\,\bigr|\,\widetilde{S}\bigr)}
    {\frac{\partial}{\partial v} G\bigl(v,
    \widetilde{S},u\bigr)\bigl|_{v=G^{-1}_{(\widetilde{S},u)}(q_\alpha(u))}}\right]\\
    & & \quad -\,\mathrm{E}\left[
    \frac{g_i\bigl(G^{-1}_{(\widetilde{S},u)}(q_\alpha(u)),
    \widetilde{S}\bigr)
    h\bigl(G_{(\widetilde{S}, u)}^{-1}(q_\alpha(u))\,\bigr|\,\widetilde{S}\bigr)}
    {\frac{\partial}{\partial v} G\bigl(v,
    \widetilde{S},u\bigr)\bigl|_{v=G^{-1}_{(\widetilde{S},u)}(q_\alpha(u))}}\right]\Bigg\}\\
    & = & 0,
\end{eqnarray*}
by Theorem \ref{pr:3}. By means of \eqref{eq:insert} this implies the assertion.\hfill $\Box$
\begin{remark}
At first sight, formulae \eqref{eq:cond_exp} and
\eqref{eq:ES_deriv} look very much like corresponding formulae for
the derivatives of VaR and ES in \citet{GLS00}, \citet{L99}, and
\citet{Tasche99}. Note, however, that those formulae were not
derived in an asymptotic multi-factor setting like the ones here.
On the other hand, the validity of the results by \citet{GLS00},
\citet{L99}, and \citet{Tasche99} is not restricted to the case of
bounded loss variables of the assets. Therefore, the results
from this paper and the earlier results complement each other.
\end{remark}

\section{Defining a diversification measure}
\label{se:defining}

During the last few years, with regard to applications, three
properties of risk measures $\rho$ turned out to be potentially
most important:
\begin{itemize}
    \item \textbf{Positive homogeneity.} See Assumption \ref{as:euler} for the
    formal definition and some comments.
    \item \textbf{Sub-additivity.} \citet[][Axiom S]{ADEH99} described
    sub-additivity
    with ``a merger does not create extra risk''. VaR as a risk measure is mainly
    criticized for lacking this property. \citet{Kalkbreneretal04} pointed out that, in the
    context of bank-internal credit risk management, the following characterization of
    sub-additivity is more apposite.
    Sub-additivity of a positively homogeneous risk measures
    is equivalent to the property that risk contributions are not larger than the corresponding
    stand-alone risks. Speaking in terms of Sub-section \ref{se:euler},
    \begin{equation}\label{eq:sub}
    \rho(V+W) \le \rho(V) + \rho(W)
    \ \text{for all}\ V,W \ \Leftrightarrow \
    \rho(V\,|\,W) \le \rho(V)\ \text{for all}\ V,W,
\end{equation}
if $\rho$ is positively homogeneous \citep[][Proposition
2.5]{Tasche02}.
    \item \textbf{Comonotonic additivity.} In actuarial science, the concept of comonotonicity
    is well-known as it supports easy and reasonably conservative representations of dependence structures
    \citep[see, e.g.,][]{Dhaeneetal}.
    Random variables $V$ and $W$ are called
    comonotonic if they can be represented as non-decreasing functions of a third
    random variable $Z$, i.e.
    \begin{subequations}
    \begin{equation}\label{eq:comono}
    V = h_V(Z) \quad \text{and}\quad W = h_W(Z)
\end{equation}
for some non-decreasing functions $h_V, h_W$. As comonotonicity is
implied if $V$ and $W$ are correlated with correlation coefficient
1, it generalizes the concept of linear dependence. A risk measure
$\rho$ is called comonotonic additive if for any comonotonic
random variables $V$ and $W$
\begin{equation}\label{eq:add}
    \rho(V+W)\ = \ \rho(V) + \rho(W).
\end{equation}
\end{subequations}
Thus comonotonic additivity can be interpreted as a specification of the worst case scenarios
for the sub-additivity \eqref{eq:sub}: nothing worse can occur than comonotonic random variables --
which seems quite natural.
\end{itemize}
Note that VaR is positively homogeneous and comonotonic additive but not sub-additive and that
ES is positively homogeneous, comonotonic additive and sub-additive \citep[see, e.g.][]{Tasche02}.
As a consequence, finding worst case scenarios for given marginal distributions of $V, W$ in \eqref{eq:sub}
is easy in case of ES (take the comonotonic scenario) and non-trivial in case of VaR
\citep[see][]{Embrechtsetal, Luciano01}.

As for positively homogeneous, comonotonic additive and sub-additive risk measures nothing
worse than the comonotonic case can happen, it seems natural to measure diversification
by comparison with the comonotonic scenario\footnote{%
\citet{MartinTasche} suggest another approach to measuring diversification as
they calculate the proportions of systematic and idiosyncratic risk
within the total risk of the portfolio.
}. This suggests the following definition.
\begin{definition}\label{de:factor}
Let $X_1, \ldots, X_n$ be real-valued random variables and let $Y
= \sum_{i=1}^n X_i$. If $\rho$ is a risk measure such that
$\rho(Y), \rho(X_1), \ldots, \rho(X_n)$ are defined, then
\begin{equation*}
    \mathrm{DF}_{\rho}(Y) \ = \
    \frac{\rho(Y)}
    {\sum_{i=1}^n \rho(X_i)}
\end{equation*}
denotes the \emph{diversification factor} of portfolio $Y$ with
respect to
the risk measure $\rho$.\\
The fraction
\begin{equation*}
    \mathrm{DF}_{\rho}(X_i\,|\,Y) \ = \
    \frac{\rho(X_i\,|\,Y)}{\rho(X_i)}
\end{equation*}
with $\rho(X_i\,|\,Y)$ being the risk contribution of $X_i$ as in
Definition \ref{de:risk_contrib} denotes the \emph{marginal
diversification factor} of sub-portfolio
$X_i$ with respect to the risk measure $\rho$.
\end{definition}
Note that without calling the concept ``diversification factor'',
\citet{MemmelWehn} calculate a diversification factor for the
German supervisor's market price risk portfolio. \citet{Garcia04} use the
diversification factors as defined here
for a representation of portfolio risk as a ``diversification
factor''-weighted sum of stand-alone risks.

If $\rho$ is sub-additive and positively homogeneous, then by
\eqref{eq:sub} both $\mathrm{DF}_{\rho}(Y)$ and
$\mathrm{DF}_{\rho}(X_i\,|\,Y)$ will be bounded by 1. If $\rho$
is additionally comonotonic additive, then the bound 1 can be
reached by portfolios with comonotonic risks. Thus, with a
reasonable risk measure, $\mathrm{DF}_{\rho}(Y)$ being close to 1
will indicate that there is no significant diversification in the
portfolio. Similarly, a value of $\mathrm{DF}_{\rho}(X_i\,|\,Y)$
close to 1 will indicate that there is almost no diversification
effect with asset $i$. As the dependence -- measured as degree of
comonotonicity -- in a portfolio is influenced both by the
idiosyncratic and the systematic risk factors, the diversification
factors according to Definition \ref{de:factor} capture name
diversification as well as sectoral diversification.

Although VaR is not a sub-additive risk measure in general,
some authors argue that, by practical experience, it can be considered an almost
sub-additive risk measure \citep[cf.][]{Danielssonetal}.
This
observation, however, seems to be strongly dependent on the context.
For instance, \citet[][Section 2.3]{FreyMcNeil} present an example of a credit
portfolio where
``measuring risk with VaR can lead to nonsensical
results''. In the
following section we illustrate the use of the
diversification factors from Definition \ref{de:factor} by a
numerical example. We consider VaR as the underlying risk measure
because this facilitates the calculations.

\section{Numerical example}
\label{se:example}
\setcounter{figure}{0}

In this section, we illustrate the application of the formulae for
the loss distribution function (Proposition \ref{pr:2}) and the
risk contributions to VaR (Theorem \ref{pr:3}) with a simple
example. We consider a special case of model \eqref{eq:new_model}
with two normally distributed systematic factors and normally
distributed idiosyncratic risk drivers as in Example
\ref{ex:idio_phi}.

As for \eqref{eq:normal_xi}, we consider the case $n=2$ which may
be interpreted as having a portfolio with a large number of credit
instruments in two different sectors. Both of these sectors are
exposed to the first systematic factor, but only the first sector
is also exposed to the second factor. By varying the extent of
this exposure to the second systematic factor we will obtain a
picture of the effect of sectoral diversification by
dependence on more than one systematic factors. Additionally, we
will fix the exposure to the second systematic factor but vary the
weights of the sectors within the portfolio in order to get an
impression of the influence of the sectoral structure on the diversification factors defined
in Section \ref{se:defining}.
\begin{example}\label{ex:num}
Consider the loss variable $\hat{L}(u)$ from Example
\ref{ex:idio_phi} in the case $n=2$ with standard normally
distributed systematic factors $S_1$ and $S_2$ and independent
(also of $(S_1, S_2)$) standard normally distributed idiosyncratic
risk drivers $\xi_1,\xi_2$. We consider the case of relative (to
the total exposure) loss, i.e.\ the case $u_1+u_2 = 1$. Hence,
$\hat{L}(u)$ reads here
\begin{equation}\label{eq:exam}
    \hat{L}(u) \ = \ u\,\Phi\bigg(\frac{t_1 - \sqrt{\varrho_{1,1}}\,S_1 -
     \sqrt{\varrho_{1,2}}\,S_2}{\omega_1}\bigg) + (1-u)\,
     \Phi\bigg(\frac{\Phi^{-1}(p) - \sqrt{\varrho_{2,1}}\,S_1- \sqrt{\varrho_{2,2}}\,S_2}
     {\omega_2}\bigg).
\end{equation}
With respect to the correlations
with the systematic factors, we fix some $\varrho \in (0,1)$ and let
\begin{subequations}
\begin{alignat}{3}\label{eq:rhos}
\varrho_{1,1} &= \sqrt{\frac{\varrho\,w}
{1+2\,w\,(1-w)\,\tau}},\quad & \varrho_{1,2}&=
\sqrt{\frac{\varrho\,(1-w)}
{1+2\,w\,(1-w)\,\tau}},\\
    \varrho_{2,1} &= \sqrt{\varrho},
    &\varrho_{2,2} &=0,
    \label{eq:rhos2}
\end{alignat}
where $w\in[0,1]$ is a weight parameter controlling the exposure
of the first sector to the second factor and $\tau$ denotes the
correlation of $S_1$ and $S_2$, i.e.\
$\tau=\mathrm{corr}[S_1,\,S_2]$. Choosing the $\varrho_{i,j}$ as
in \eqref{eq:rhos} and \eqref{eq:rhos2} implies by
\eqref{eq:omegas}
\begin{equation}\label{eq:omega}
    \omega_1\ =\ \sqrt{1-\varrho}\ =\ \omega_2.
\end{equation}
\end{subequations}
\end{example}
The square-root representation in \eqref{eq:rhos} and
\eqref{eq:rhos2} was chosen in order to make the correlations
comparable in size with those from \citet[][\S 272]{BC04}. By
choosing $\varrho_{1,1}$ and $\varrho_{1,2}$ as in
\eqref{eq:rhos}, we ensure that in \eqref{eq:exam} both factor
combinations $\varrho_{i,1}\,S_1 + \varrho_{i,2}\,S_2$, $i=1,2$,
are identically normally distributed with variance $\varrho$. As a
consequence, the weights on idiosyncratic risk are the same across the
two sectors. Thus, in the example
all observed differences in VaR or in the risk contributions are
due to the sector structure only. A weight $w=1$ means that we are
in a single factor model, whereas $w=0$ implies that the factors
that drive the risk of the both sectors  have correlation $\tau$
and are, in particular, independent for $\tau=0$.

For the first calculations we choose
\begin{equation}\label{eq:values}
    t_1\ = \ \Phi^{-1}(0.1)\ =\ t_2, \quad \varrho \ = \ 0.1, \quad u\ = \ 0.1.
\end{equation}
This choice is mainly driven by the desire to come up with
illustrative results. The value $10\%$ for $\varrho$ is somewhere
in the center of the span provided by \citet{BC04}. The choice for
the threshold values $t_1$ and $t_2$ may be interpreted as having
a large credit portfolio with two sectors, both with an average
probability of default of 10\%.
  \begin{table}[htbp]
    \begin{center}
\parbox{\captionwidth}{\caption{\label{tab:1}\emph{VaRs at different confidence levels
$\alpha$ for the asymptotic single ($w=1$) and two-factor ($w=0$)
models as in Example
\ref{ex:num}. Parameter values as in \eqref{eq:values}. Independent systematic
factors ($\tau=0$).}}}\\[2ex]
\begin{tabular}{c||c|c|c|c|c|c|c}
$\alpha$ & 75\% & 90\% & 95\% & 97.5\% & 99\% & 99.9\% & 99.95\%
\\ \hline VaR (single factor) & 13.0\% & 17.8\% & 21.1\% & 24.3\%
& 28.3\% & 37.4\% & 40.0\%\\ \hline VaR (two factors) & 12.7\% &
17.0\% & 20.0\% & 22.9\% & 26.5\% & 34.7\% & 37.0\%\\ \hline Ratio
of the above & 97.9\% & 95.8\% & 94.9\% & 94.3\% & 93.7\% & 92.8\%
& 92.6\%
\end{tabular}
    \end{center}
  \end{table}

In order to assess the impact on sectoral diversification by
several systematic factors at portfolio level,
first we calculate\footnote{%
The calculations for the following examples require numerical
root-finding and integration. Files with the R-code
\citep[cf.][]{R03} used by the author can be provided upon
request.%
} VaR-figures at different confidence levels both for the single factor model
as in \eqref{eq:exam} with $w=1$ as well as for the two-factor
model with $w=0$. Table \ref{tab:1} for independent systematic
factors and Table \ref{tab:2} for positively correlated systematic
factors
 show that the impact even in the case of an
independent second factor and for high levels of VaR remains
limited. The results from Table \ref{tab:2}, compared to those of
Table \ref{tab:1}, reflect the non-surprising fact that, with
positively correlated systematic factors, the potential for
diversification is less than in the case of independent systematic
factors.
\begin{table}[htbp]
    \begin{center}
\parbox{\captionwidth}{\caption{\label{tab:2}\emph{VaRs at different levels
$\alpha$ for the asymptotic single ($w=1$) and two-factor ($w=0$)
models as in Example
\ref{ex:num}. Parameter values as in \eqref{eq:values}. Systematic
factors with 50\% correlation ($\tau=0.5$).}}}\\[2ex]
\begin{tabular}{c||c|c|c|c|c|c|c}
$\alpha$ & 75\% & 90\% & 95\% & 97.5\% & 99\% & 99.9\% & 99.95\%
\\ \hline
VaR (single factor) & 13.0\% & 17.8\% & 21.1\% & 24.3\% & 28.3\% & 37.4\% & 40.0\%\\
\hline VaR (two factors) & 12.9\% & 17.4\% & 20.6\% & 23.6\% & 27.4\% & 36.1\% & 38.5\%\\ \hline
Ratio of the above & 99.2\% & 98.1\% & 97.6\% & 97.3\% & 96.9\% & 96.4\% & 96.3\%
\end{tabular}
    \end{center}
  \end{table}

The impact of sectoral diversification turns out to be much higher
if we consider UL contributions with respect to VaR instead of
total VaR. ``UL'' means ``unexpected loss'' and is defined by
choosing
\begin{equation}\label{eq:UL}
    \rho(V)\ =\ \mathrm{VaR}_\alpha(V) - \mathrm{E}[V]\ =\
    q_\alpha(V) - \mathrm{E}[V]\ = \ \mathrm{UL}(V)
\end{equation}
in Definition \ref{de:risk_contrib}. In Figure \ref{fig:1} we plot
the relative contribution to UL with respect to 99.9\%-VaR of the
first sector in the model in Example \ref{ex:num} (i.e.\
the ratio of the contribution of the first sector
to UL in the sense of Definition \ref{de:risk_contrib} and portfolio-wide
UL) against the
extent of the sector's exposure to the first factor (low values of
$w$ correspond to low exposure, values of $w$ close to 1
correspond to high exposure). For calculating the contributions,
we applied \eqref{eq:partial_deriv}. In the case of independent
systematic factors, it turns out that the size of the risk
contribution of the first sector can be reduced to almost 0 when
it is exposed to the second systematic factor only. The rate of
the reduction becomes the smaller the stronger the exposure to the
first systematic factor but remains significant. These effects are
significantly weaker, if there is a positive, but less than 100\%
correlation of the systematic factors. There is no reduction at
all of the risk contribution, if the correlation of the systematic
factors is 100\% -- this corresponds to the case of the asymptotic
single risk factor model that underlies the Basel II risk capital charges.
\begin{samepage}
\refstepcounter{figure}
\begin{figure}[ht]
\centering
  \parbox{14.0cm}{Figure \thefigure:
  \emph{Relative contribution to UL with respect to 99.9\%-Var of the first sector
in the model in Example \ref{ex:num} as function of the extent of
the sector's exposure  (measured by $w\in[0,1]$) to the first
factor. Parameter values as in \eqref{eq:values}. Factor
correlations 50\%, 0\%, and 100\% respectively.}}
\label{fig:1}\\[2ex]
\ifpdf
    \resizebox{\height}{10.0cm}{\includegraphics[width=16.0cm]{risk_contribution.pdf}}
\else
\begin{turn}{270}
\resizebox{\height}{16.0cm}{\includegraphics[width=10.0cm]{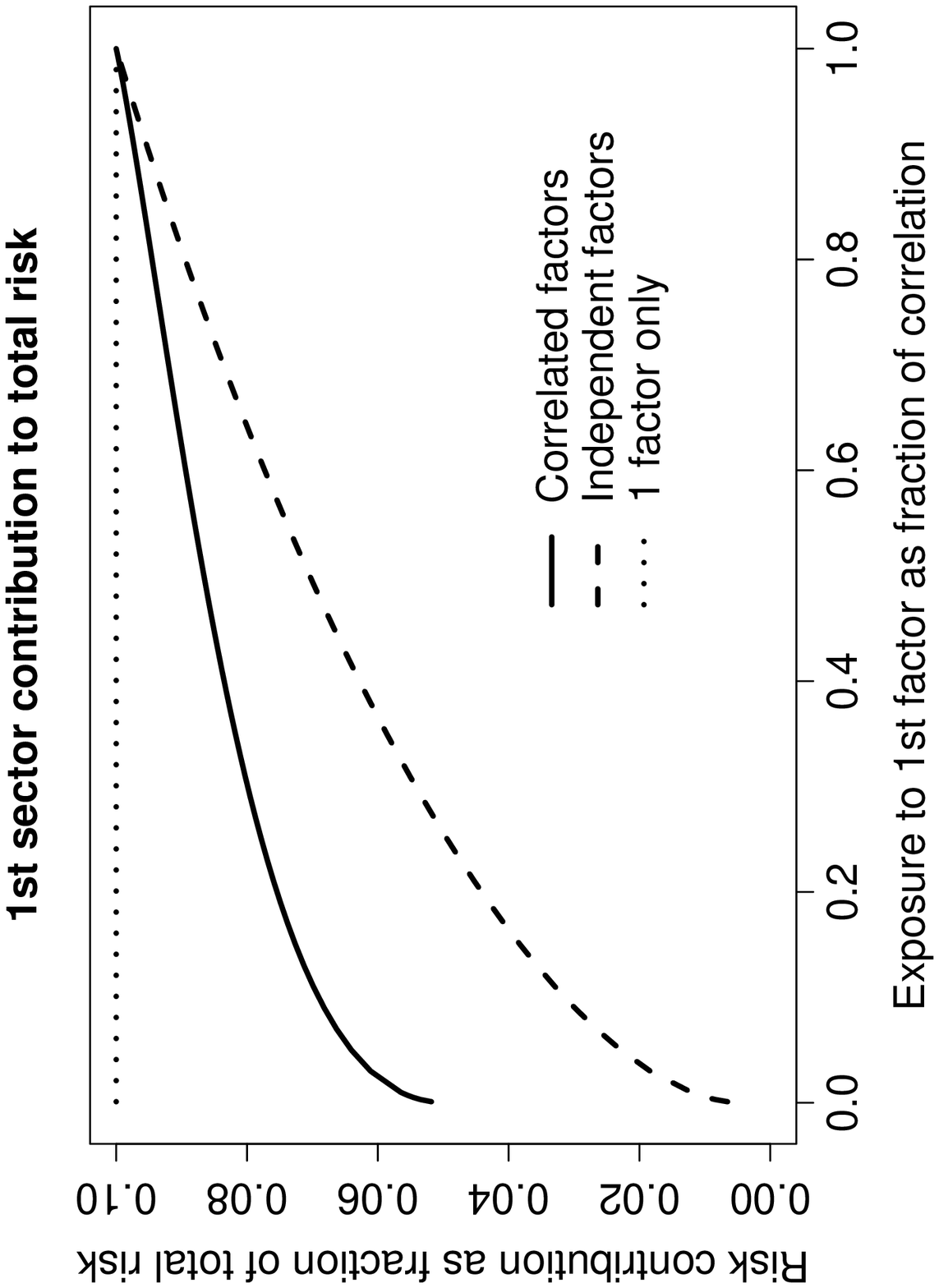}}
\end{turn}
\fi
\end{figure}
\end{samepage}

In order to illustrate the functioning of the diversification
factors defined in Section \ref{se:defining} we fix the exposure
of the first sector to the second systematic factor by setting $w
= 0.5$. We then make the weight $u$ (see \eqref{eq:exam}) of the
first sector in the portfolio move from 0\% to 100\%. As in the
case of Figure \ref{fig:1}, we calculate with two different values
for the correlation of the two systematic factors. The
parameter settings for Figure \ref{fig:2} are
given in Equations \eqref{eq:new_parameters1} and
\eqref{eq:new_parameters2} respectively.
\begin{subequations}
\begin{alignat}{5}\label{eq:new_parameters1}
  \tau &= 0, &\quad t_1 = \Phi^{-1}(0.2), &\quad t_2 = \Phi^{-1}(0.1), &\quad\varrho&=0.1 \\
  \tau &= 0.5, &\quad t_1 = \Phi^{-1}(0.2), &\quad t_2 = \Phi^{-1}(0.1), &\quad\varrho&=0.1.
  \label{eq:new_parameters2}
\end{alignat}
\end{subequations}
Figure \ref{fig:2}  illustrates the connection
between the two types of diversification factors from Definition
\ref{de:factor}. In both panels, the solid line  shows the
portfolio-wide diversification factor of the loss variable from
\eqref{eq:exam}. The dashed lines in the panels of Figure \ref{fig:2}
reflect the corresponding marginal diversification
factor of the first sector, whereas the dotted lines give the
marginal diversification factor of the second sector. The first panel of Figure
\ref{fig:2} represents the case of independent systematic factors, whereas
the second panel of Figure \ref{fig:2} shows the case of positively correlated
systematic factors.
From the definition of the risk contribution by means of a
derivative (see Definition \ref{de:risk_contrib}), in both cases
it follows that the three lines intersect at just the weight of
the first sector that yields the most diversified portfolio in the
sense of being the portfolio with the minimum diversification
factor to be feasible by changing the sector weights.
\refstepcounter{figure}
%
\begin{center}
  \parbox{14.0cm}{Figure \thefigure:
    \emph{Diversification factors with respect to UL in sense of Definition \ref{de:factor}
  for loss variable
  $\hat{L}(u)$ and sectors from Example \ref{ex:num}, with $w=0.5$. Represented as functions of
  weight $u\in [0,1]$ of first sector. First panel for case of
 independent systematic factors, second panel for case of correlated systematic factors.
 Parameter setting specified by \eqref{eq:new_parameters1} and
  \eqref{eq:new_parameters2} respectively.}
  }%
\label{fig:2}\\[2ex]
\ifpdf
\resizebox{\height}{15.0cm}{\includegraphics[width=12.0cm]{div_index_partial.pdf}}
\else
\begin{turn}{270}
\resizebox{\height}{12.0cm}{\includegraphics[width=15.0cm]{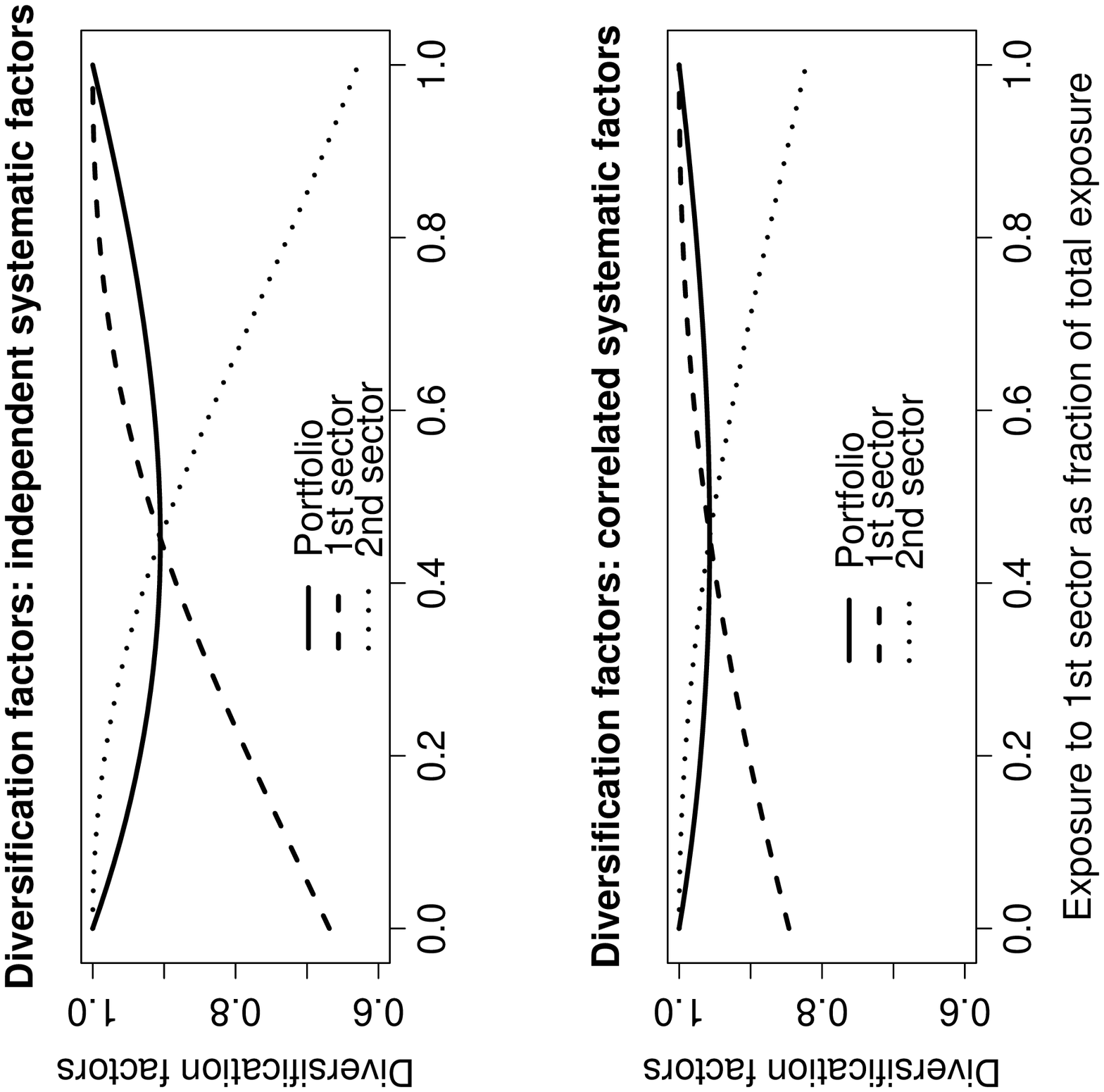}}
\end{turn}
\fi
\end{center}

According to Figure \ref{fig:2},
 those portfolios where
portfolio-wide and marginal diversification factors are close
together could be called well-diversified. A wide range of the
diversification factors of a portfolio would then indicate that
the portfolio is not very well diversified.
However, Figure \ref{fig:2}  demonstrates that the
possible range of the diversification factors depends on the
potential for diversification. The smaller maximum possible range
of the marginal diversification in the correlated case correctly
indicates that in the extreme case of exposure to only one
systematic factor the portfolio is closer to the optimum than it
is in the independent case. But this is only due to the fact that
in the case of positively correlated systematic factors, the
potential for diversification is less than in the case of
independent systematic factors, as follows from the higher value
of the minimum portfolio-wide diversification factor.

\section{Conclusions}
\label{se:concl}
In this paper we have derived closed-form formulae for risk
contributions
    to VaR and ES in the context of asymptotic multi-factor models,
    thus generalizing the capital requirements as provided
    by the Basel II Accord in the context of the ASRF (Asymptotic Single
    Risk Factor) model. The effort needed for the numerical calculations
    is higher than in the ASRF case but, as a numerical example shows,
    remains feasible at least in the case of two-factor models.
    The example also indicates that the effect of sectoral diversification
    by several systematic factors
    on portfolio-wide economic capital is moderate but can
    be significant for risk contributions of single assets, sectors or sub-portfolios.

The risk contributions we have analyzed in the first sections of
the paper can be used for calculating diversification factors for
sub-portfolios, sectors or assets in a portfolio. If these
factors, considered for all the assets, sectors of sub-portfolios
in the portfolio, take a wide range, then there is a high
potential for diversification in the portfolio. If, in contrast,
the range of the factors is narrow, there is not much potential
left for diversification by changing the weights of the assets,
sectors or sub-portfolios in the portfolio. In this case, more
diversification can only be reached by adding new assets or by
removing assets from the portfolio.

 This observation suggests the use of the newly developed diversification factors
   for reflecting
    sectoral diversification by several systematic factors: assets, sectors or sub-portfolios
    found well-diversified by a
    marginal diversification
    factor close to
    the portfolio-wide diversification factor
    could receive a reduction of capital requirements. The sizes of such
    reductions could be estimated by means of an asymptotic multi-factor
    model. Of course, the concrete choice of the model and its underlying
    parameters might have a strong impact on the estimates. Further research
    in this direction seems necessary.


%
\end{document}